\documentclass[aps,twocolumn,showpacs,amsmath,amssymb,superscriptaddress,prc]{revtex4-2}
\usepackage{graphicx,here}
\usepackage{multirow}
\usepackage{tikz}
\usepackage{epstopdf}
\usepackage[percent]{overpic}
\usepackage{scrextend}
\usepackage{xcolor,xspace}
\usepackage[ulem=normalem]{changes}
\usepackage{hyperref}
\hypersetup{colorlinks=True,urlcolor=blue,linkcolor=blue,citecolor=blue,filecolor=black}

\colorlet{Changes@Color}{red}
\usepackage{dcolumn,color,footnote,bm}
\usepackage{url,longtable,tabularx}
\usepackage{threeparttable}
\usepackage{siunitx}
\usepackage{physics}

\usepackage{fancybox}
\usetikzlibrary{matrix}

\newcommand\+{\dagger}
\newcommand\jr{j_{\rho}}
\newcommand\mr{m_{\rho}}

\newcommand\mgt{M_{\mathrm{GT}}}
\newcommand\mf{M_{\mathrm{F}}}

\newcommand\hb{\hat{H}_{\mathrm{B}}}
\newcommand\hf{\hat{H}_{\mathrm{F}}}
\newcommand\hbf{\hat{V}_{\mathrm{BF}}}
\newcommand\hff{\hat{V}_{\nu\pi}}

\newcommand\ga{g_{\mathrm{A}}}
\newcommand\gv{g_{\mathrm{V}}}

\newcommand\ft{\log{}_{10}ft}
\newcommand\btm{\beta^{-}}
\newcommand\btp{\beta^{+}}

\newcommand\vd{v_{\mathrm{d}}}

\newcommand\vsss{v_{\mathrm{ss}}}
\newcommand\vt{v_{\mathrm{t}}}

\begin{document}

\title{Sensitivity analysis of    
$\beta$-decay half-life predictions for Ge, As, Zr and Mo nuclei 
within the mapped interacting boson model}

\author{M. Homma}
\affiliation{Department of Physics, 
Hokkaido University, Sapporo 060-0810, Japan}

\author{K. Nomura}
\email{nomura@sci.hokudai.ac.jp}
\affiliation{Department of Physics, 
Hokkaido University, Sapporo 060-0810, Japan}
\affiliation{Nuclear Reaction Data Center, 
Hokkaido University, Sapporo 060-0810, Japan}

\date{\today}

\begin{abstract}
We analyze parameter sensitivities of the 
mapped interacting boson model (IBM) and 
boson-fermion-fermion model (IBFFM) in the 
description of $\beta$-decay properties 
of the even-mass neutron-deficient Ge and As, 
and neutron-rich Zr and Mo isotopes. 
Based on the 
self-consistent mean-field calculations with a 
given energy density functional and a pairing interaction, 
the IBM Hamiltonian for even-even nuclei, 
single-particle energies, 
and occupation probabilities for unpaired nucleons, 
which are necessary building blocks of 
the IBFFM Hamiltonian and Gamow-Teller 
and Fermi transition operators, 
are completely determined. 
A few coupling constants of the boson-fermion and 
residual neutron-proton interactions are only 
phenomenological parameters fitted to reproduce 
low-energy spectra of odd-mass and odd-odd nuclei.
It is found that the calculated $\log{}_{10}ft$ 
values for the $\beta^+$ decays $^{68}$As$\to^{68}$Ge 
are particularly sensitive to the 
quadrupole-quadrupole boson interaction strength 
used for the parent ($^{68}$As) nucleus. 
We further incorporate higher-order terms in the 
one-nucleon transfer operators in the boson system, 
and find that, while their effects are non-negligible, 
they do not significantly alter qualitative 
features of $\beta$-decay properties. 
We report a novel application of the mapped IBM framework 
to compute $\beta$-decay half-lives, and show that the 
observed trend along isotopic chains 
are reasonably reproduced. 
\end{abstract}

\maketitle

\section{Introduction}

The nuclear $\beta$ decay plays a major role 
in the synthesis of heavy chemical elements 
in the early universe, and is thus of fundamental 
importance in nuclear physics and related fields. 
While light elements such as hydrogen, helium, and lithium, 
which have masses $A \le 7$, are created in 
big-bang nucleosynthesis, and heavier elements up 
to iron are synthesized as stars evolve, 
elements heavier than iron 
are considered to be produced by the $s$ and $r$ processes, 
which proceed via neutron capture and $\beta$ decay. 
The $s$ process follows the $\beta$-stability line, 
and plenty of experimental data are available for the stable nuclei. 
On the other hand, 
the $r$ process follows a path on the neutron-rich side, 
which is far from the stability line, 
and it is even more challenging to study neutron-rich 
heavy nuclei experimentally 
\cite{dillmann2003,nishimura2011,quinn2012,lorusso2015,caballero2016}. 
Therefore, 
current studies of the $r$ process rely heavily 
on theoretical predictions on 
physical quantities such as the neutron capture 
cross sections and $\beta$-decay half-lives.

It is thus crucial to develop theoretical models 
that accurately describe characteristics of $\beta$ decay, 
and to reduce or make controllable the 
possible theoretical uncertainties in given models. 
Nuclear structure models that are frequently applied to study 
the $\beta$-decay of medium-heavy and heavy nuclei include 
the nuclear shell model (NSM) 
\cite{langanke2003,caurier2005,syoshida2018,suzuki2018,kumar2024}, 
the quasiparticle random-phase approximation (QRPA) 
\cite{moeller2003,alvarez2004,sarriguren2015,boillos2015,pirinen2015,simkovic2013,mustonen2016,marketin2016,suhonen2017,ney2020,ravlic2021,yoshida2023}, 
and the interacting boson model (IBM) \cite{navratil1988,dellagiacoma1989,brant2004,brant2006,yoshida2013,mardones2016,nomura2020beta-1,nomura2020beta-2,ferretti2020,nomura2022beta-ge,nomura2022beta-rh,nomura2024beta,homma2024}. 
The basic assumption of 
the IBM is that collective monopole 
(with spin and parity $J^{\pi}=0^+$)
and quadrupole ($J^{\pi}=2^+$) pairs of valence nucleons 
are approximated to $s$ and $d$ bosons, respectively \cite{IBM}, 
which therefore represents a drastic truncation 
of the full shell-model configuration space. 
The IBM has been often employed in the $\beta$-decay 
studies in different mass regions. 
While the IBM is able to simultaneously 
describe low-lying collective states and $\beta$ decays
of heavy nuclei in deformed or open-shell regions, 
in the majority of the earlier $\beta$-decay studies 
within the IBM most of the model parameters 
were obtained from phenomenological adjustments 
to the experimental low-energy spectra for each nucleus. 
The reliability of 
the model in the predictions of $\beta$-decay properties 
of neutron-rich nuclei that are very far from the 
stability line and for which experimental data are 
not available has not yet been clarified.

In recent years 
a method of calculating $\beta$-decay properties 
has been developed  
\cite{nomura2020beta-1,nomura2020beta-2,nomura2022beta-ge,nomura2024beta} 
that consists of mapping the 
self-consistent mean-field (SCMF) solutions 
onto the IBM \cite{nomura2008,nomura2010,nomura2011rot}. 
In that method, strength parameters of the 
IBM Hamiltonian describing even-even nuclei 
are determined microscopically so that the potential energy 
surface (PES) obtained from the self-consistent mean-field (SCMF) 
calculation employing a given energy 
density functional (EDF) and a pairing force 
is mapped onto 
the expectation value of the boson Hamiltonian. 
Odd-odd nuclear systems are described 
in terms of the 
interacting boson-fermion-fermion model 
(IBFFM) \cite{IBFM,brant1984}, which consists of 
the even-even (IBM) core representing the 
collective motion, and an unpaired neutron 
and an unpaired proton as single-particle 
degrees of freedom. 
The SCMF calculation provides single-particle 
energies and occupation probabilities for odd nucleons, which 
are used as a microscopic input to build the IBFFM Hamiltonian. 
However, only the strength parameters 
for the interactions between unpaired nucleons and even-even boson 
cores and 
between unpaired nucleons are determined 
so as to reproduce the low-lying energy spectra 
of each odd-odd nucleus 
to a certain accuracy. 
The wave functions of the IBM and IBFFM, 
which give reasonable 
descriptions of low-lying states of even-even and odd-odd nuclei, 
respectively, are used to compute Gamow-Teller (GT)
and Fermi transition strengths for 
the $\beta$ decays of interest.

In Ref.~\cite{nomura2024beta}, 
in particular, the mapped IBM was 
applied to study $\beta$-decay properties 
as well as low-lying collective 
states of even-even and odd-odd nuclei from Kr to Cd isotopes 
near $N=60$, using the relativistic EDF 
as a microscopic input. 
In that study, 
the observed systematic of the $\beta$-decay 
$\ft$ values along isotopic chains was reproduced 
reasonably well, but 
some of the measured $\ft$ values for 
those nuclei with $N\leqslant 60$ 
were significantly underestimated. 
To address the deficiency, 
in a subsequent paper \cite{homma2024}
we studied the parameter dependence of the predicted $\ft$ values 
for the $\btm$ decay of the Zr isotopes within 
the EDF-mapped IBM framework. 
By varying IBFFM as well as IBM parameters used 
in Ref.~\cite{nomura2024beta}, we found that the $\ft$ values 
for the $\btm$ decay of Zr isotopes depended strongly on 
the quadrupole-quadrupole boson interaction strength and 
that of the residual neutron-proton 
interaction of tensor type in the IBFFM 
Hamiltonian for the daughter (Nb) nuclei. 
Furthermore, it was found that, when 
these parameters were adjusted to the experimental 
$\ft$ values, they led to a better reproduction of 
the energy spectra of the Nb isotopes and their boson 
core (Mo) nuclei.

In the present study, we extend the analysis of Ref.~\cite{homma2024} 
to the $\beta$ decay of neutron-deficient Ge and As nuclei 
in the mass $A\approx 70$ region to see if the conclusion 
drawn in Ref.~\cite{homma2024} is valid in other mass regions. 
As a representative case, 
we specifically consider 
the $\btp$ decay of $^{68}$As. 
We also incorporate higher-order terms in the 
one-nucleon transfer operators in the IBFFM in constructing 
the GT and Fermi transition operators, 
and study the role played by these terms in 
the corresponding matrix elements for the $\btp$ decays of 
$^{68,70}$As, and $\btm$ decays of 
$^{78}$Ge, and even-even $^{98-110}$Zr 
and $^{102-110}$Mo isotopes. 
Here, the choice of the specific regions 
$A\approx 70$ and $A\approx 100$ 
is motivated by fact that they are 
rather close to  $^{76}$Ge, $^{96}$Zr, and $^{100}$Mo, 
which are candidate nuclei for the neutrinoless 
double-$\beta$ decay, a fundamental nuclear decay 
under extensive investigations \cite{avignone2008,engel2017,agostini2023}. 

We further report a novel application of the mapped IBM 
framework to compute $\beta$-decay half-lives 
of the considered Ge, As, Zr, and Mo isotopes. 
While the $\beta$-decay half-lives are extensively studied 
by various theoretical approaches 
such as NSM and QRPA, the IBM has mostly been applied to 
compute the GT and Fermi transitions to only a few low-lying states. 
An exception is perhaps the IBM-2 calculations 
of the double-$\beta$ decay nuclear matrix elements 
in Refs.~\cite{yoshida2013,nomura2022bb,nomura2024bb}, 
where a number of intermediate states of odd-odd nuclei 
were computed without the closure approximation. 
To the best of our knowledge, the application of the 
IBM to systematic half-life studies 
for neutron-rich isotopes has rarely been attempted.

The paper is organized as follows. 
Section~\ref{sec:model} provides a brief reminder of the 
EDF-mapped IBM and IBFFM methods, and calculated 
results for low-energy spectra of 
even-mass nuclei. 
In Sec.~\ref{sec:beta} $\beta$-decay properties are 
discussed, including the parameter sensitivity 
of the $^{68}$As $\btp$ decay, 
effects of the higher-order terms, 
and predictions on the half-lives. 
Section~\ref{sec:summary} gives a summary of the principal 
results and perspectives for future studies.

\section{Mapped IBM-2 and IBFFM-2\label{sec:model}}

This section describes the procedure to construct 
the IBFFM Hamiltonian from the EDF calculation, 
gives definitions of the GT and Fermi operators, 
and presents calculated results for low-energy spectra.

\subsection{Model Hamiltonian\label{sec:ham}}

\begin{table}
\caption{
Configurations of the odd-odd nuclei under study (first column) 
with their neutron numbers $N$ (second column). 
Corresponding even-even boson core nuclei are shown 
in the third column. 
The fourth (fifth) column shows whether the odd neutron 
(proton) is of particle ($p$) or hole ($h$) nature.}
    \label{tab:core}
    \begin{ruledtabular}
    \begin{tabular}{lcccc}
Odd-odd & $N$ & Even-even core & Neutron & Proton \\
\hline
$_{33}$As$_{N}$ & $35 \le N \le 39$ & $_{32}$Ge$_{N-1}$ & $p$ & $p$ \\
$_{33}$As$_{N}$ & $41 \le N \le 45$ & $_{32}$Ge$_{N+1}$ & $h$ & $p$ \\
$_{41}$Nb$_{N}$ & $55 \le N \le 65$ & $_{42}$Mo$_{N-1}$ & $p$ & $h$ \\ 
   & $N=67, 69$ & $_{42}$Mo$_{N+1}$ & $h$ & $h$ \\ 
     $_{43}$Tc$_{N}$ & $55 \le N \le 65$ & $_{44}$Ru$_{N-1}$ & $p$ & $h$ \\
                    & $N=67$ & $_{44}$Ru$_{N+1}$ & $h$ & $h$ \\
    \end{tabular}
    \end{ruledtabular}
\end{table}

As in the previous studies 
of Refs.~\cite{nomura2024beta,homma2024}, 
we use the neutron-proton IBM 
(IBM-2) and IBFFM (IBFFM-2). 
The building blocks of the IBM-2 are 
neutron $s_\nu$ and $d_\nu$ bosons, and proton 
$s_{\pi}$ and $d_\pi$ bosons. 
The $s_{\nu}$ and $d_\nu$ ($s_\pi$ and $d_\pi$) bosons 
are associated with the collective monopole and 
quadrupole pairs of valence neutrons (protons), 
respectively \cite{IBM,OAIT,OAI}. 
We adopt the IBM-2 Hamiltonian of the same 
form as used in Refs.~\cite{nomura2024beta,homma2024}: 
\begin{align}
\label{eq:bham}
 \hb = 
&\epsilon_{d}(\hat{n}_{d_{\nu}}+\hat{n}_{d_{\pi}})
+\kappa\hat{Q}_\nu \cdot\hat{Q}_\pi
\nonumber\\
&+\kappa_{\nu}\hat{Q}_\nu \cdot\hat{Q}_\nu
+\kappa_{\pi}\hat{Q}_\pi \cdot\hat{Q}_\pi
+ \kappa'\hat{L}\cdot\hat{L} \; .
\end{align}
$\hat n_{d_\rho} = d^\+_\rho\cdot\tilde d_\rho$ 
($\rho=\nu$ or $\pi$) denotes the $d$-boson number operator 
with $\tilde{d}_{\rho\mu} = (-1)^\mu d_{\rho -\mu}$, 
and $\epsilon_d$ is the single $d$ boson 
energy relative to that of $s$ bosons. 
The second, third, and fourth terms are 
the quadrupole-quadrupole 
interactions between neutron and proton bosons, 
between neutron and neutron bosons, 
and between proton and proton bosons, 
respectively. 
The quadrupole operator $\hat Q_{\rho}$ is 
defined as 
$\hat Q_{\rho} = s^\+_\rho \tilde d_\rho + d^\+ s_{\rho} + \chi_{\rho} (d^\+_\rho\times\tilde d_\rho)^{(2)}$, 
where $\chi_\nu$ and $\chi_\pi$ are 
dimensionless parameters. 
$\kappa$, $\kappa_{\nu}$, and $\kappa_\pi$ 
are strength parameters of the 
quadrupole-quadrupole interactions. 
The last term of Eq.~(\ref{eq:bham}) 
stands for a rotational term, 
with $\kappa'$ being 
the strength parameter, and 
$\hat L = \hat L_{\nu} + \hat L_{\pi}$
denotes the angular momentum operator 
with $\hat L_\rho = (d^\+_\rho\times\tilde d_\rho)^{(1)}$.

The IBFFM-2 Hamiltonian is given in general by
\begin{align}
\label{eq:ham}
 \hat{H}=\hb + \hf^{\nu} + \hf^{\pi} + \hbf^{\nu} + \hbf^{\pi} + \hff \; .
\end{align}
The first term represents the IBM-2 core 
Hamiltonian of Eq.~(\ref{eq:bham}). 
The second (third) term of 
Eq.~(\ref{eq:ham}) represents 
the single-neutron (proton) Hamiltonian of the form
\begin{align}
\label{eq:hf}
 \hf^{\rho} = -\sum_{\jr}\epsilon_{\jr}\sqrt{2\jr+1}
  (a_{\jr}^\+\times\tilde a_{\jr})^{(0)}
\equiv
\sum_{\jr}\epsilon_{\jr}\hat{n}_{\jr} \; ,
\end{align}
where $\epsilon_{\jr}$ stands for the 
single-particle energy of the odd neutron 
or proton orbital $\jr$. 
$a_{\jr}^{(\+)}$ represents 
the particle annihilation (creation) operator, 
with $\tilde{a}_{\jr}$ defined by 
$\tilde{a}_{\jr\mr}=(-1)^{\jr -\mr}a_{\jr-\mr}$. 
On the right-hand side of Eq.~(\ref{eq:hf}), 
$\hat{n}_{\jr}$ stands for the number operator 
for the odd particle. 
The single-particle space for the As isotopes taken in the 
present study comprises the neutron and proton 
$2p_{1/2}$, $2p_{3/2}$, and $1f_{5/2}$ orbitals.
The space for the Nb and Tc isotopes consists of 
the neutron $3s_{1/2}$, $2d_{3/2}$, $2d_{5/2}$, 
and $1g_{7/2}$ orbitals, and the proton $1g_{9/2}$ orbital. 
For $^{96-110}$Nb ($^{98-110}$Tc), 
since the valence neutrons and protons 
are, respectively, treated as particles 
and holes, the corresponding even-even 
boson cores are the $^{96-110}$Mo ($^{98-110}$Ru) nuclei, 
respectively. 
Configurations of the odd-odd $^{68-78}$As, $^{96-110}$Nb 
and $^{98-110}$Tc in terms of the boson core plus 
odd nucleons are summarized in Table~\ref{tab:core}.

The fourth (fifth) term 
of Eq.~(\ref{eq:ham}) denotes the interaction 
between a single neutron (or proton) and 
the even-even boson core, and is 
given as \cite{scholten1985,IBFM}
\begin{equation}
\label{eq:hbf}
 \hbf^{\rho}
=\Gamma_{\rho}\hat{V}_{\mathrm{dyn}}^{\rho}
+\Lambda_{\rho}\hat{V}_{\mathrm{exc}}^{\rho}
+A_{\rho}\hat{V}_{\mathrm{mon}}^{\rho} \; ,
\end{equation}
where the first, second, and third terms 
represent the quadrupole dynamical, exchange, 
and monopole interactions, respectively, 
with the strength parameters 
$\Gamma_\rho$, $\Lambda_\rho$, and $A_{\rho}$. 
Within the generalized seniority 
framework \cite{scholten1985,IBFM}, 
the terms in (\ref{eq:hbf}) are expressed of the forms
\begin{widetext}
\begin{align}
\label{eq:dyn}
&\hat{V}_{\mathrm{dyn}}^{\rho}
=\sum_{\jr\jr'}\gamma_{\jr\jr'}
(a^{\+}_{\jr}\times\tilde{a}_{\jr'})^{(2)}
\cdot\hat{Q}_{\rho'} \; ,
\\
\label{eq:exc}
&\hat{V}^{\rho}_{\mathrm{exc}}
=-\left(
s_{\rho'}^\+\times\tilde{d}_{\rho'}
\right)^{(2)}
\cdot
\sum_{\jr\jr'\jr''}
\sqrt{\frac{10}{N_{\rho}(2\jr+1)}}
\beta_{\jr\jr'}\beta_{\jr''\jr}
:\left[
(d_{\rho}^{\+}\times\tilde{a}_{\jr''})^{(\jr)}\times
(a_{\jr'}^{\+}\times\tilde{s}_{\rho})^{(\jr')}
\right]^{(2)}:
+ (\text{H.c.}) \; ,
\\
\label{eq:mon}
&\hat{V}_{\mathrm{mon}}^{\rho}
=\hat{n}_{d_{\rho}}\hat{n}_{\jr} \; ,
\end{align}
\end{widetext}
where the $j$-dependent factors 
$\gamma_{\jr\jr'}=(u_{\jr}u_{\jr'}-v_{\jr}v_{\jr'})Q_{\jr\jr'}$ 
and $\beta_{\jr\jr'}=(u_{\jr}v_{\jr'}+v_{\jr}u_{\jr'})Q_{\jr\jr'}$, 
with 
$Q_{\jr\jr'}=\mel*{\ell_{\rho}\frac{1}{2}\jr}{|Y^{(2)}|}{\ell'_\rho\frac{1}{2}\jr'}$ being the matrix element of the fermion 
quadrupole operator in the single-particle basis. 
$\hat{Q}_{\rho'}$ in Eq.~(\ref{eq:dyn}) denotes 
the quadrupole operator in the boson system, 
introduced in Eq.~(\ref{eq:bham}). 
The notation $:(\cdots):$ in Eq.~(\ref{eq:exc}) 
stands for normal ordering. 
Furthermore, the unperturbed single-particle 
energy, $\epsilon_{\jr}$, in Eq.~(\ref{eq:hf}) 
is replaced with the quasiparticle 
energy $\tilde\epsilon_{\jr}$.

$\hff$ in (\ref{eq:ham}) 
corresponds to the residual interaction 
between the unpaired neutron and proton. 
In this study, we consider the following form:
\begin{align}
\label{eq:hff}
\hff
=& 4\pi{\vd}
\delta(\bm{r})
\delta(\bm{r}_{\nu}-r_0)
\delta(\bm{r}_{\pi}-r_0)
\nonumber
\\
&-\frac{1}{\sqrt{3}}
\vsss \bm{\sigma}_\nu \cdot \bm{\sigma}_\pi
+ \vt
\left[
\frac{3({\bm\sigma}_{\nu}\cdot{\bf r})
({\bm\sigma}_{\pi}\cdot{\bf r})}{r^2}
-{\bm{\sigma}}_{\nu}
\cdot{\bm{\sigma}}_{\pi}
\right] \; .
\end{align}
The first, second and third terms stand for 
the $\delta$, the spin-spin, 
and tensor interactions, with 
$\vd$, $\vsss$ and $\vt$ being strength parameters, 
respectively. 
Note that $\bm{r}=\bm{r}_{\nu}-\bm{r}_{\pi}$ 
is the relative coordinate of the 
neutron and proton, and $r_0=1.2A^{1/3}$ fm. 
The matrix element of $\hff$ depends on the 
occupation $v_j$ and unoccupation $u_j$ 
amplitudes.

\subsection{Procedure to build the IBFFM-2 Hamiltonian\label{sec:set-parameters}}

The procedure to determine 
the IBFFM-2 Hamiltonian Eq.~(\ref{eq:ham}) 
consists of the following three steps.

\begin{figure}

\centering
\includegraphics[width=\linewidth]{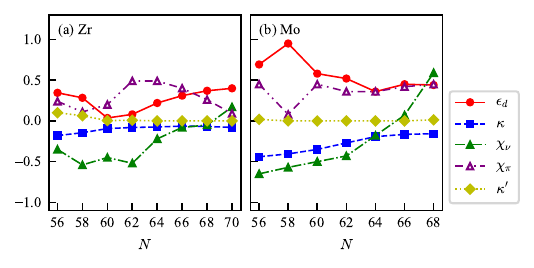}
\caption{\label{fig:para-ibm}
The IBM-2 parameters for the even-even 
(a) $^{96-110}$Zr and (b) $^{98-110}$Mo nuclei, 
which are obtained from mapping the RHB-SCMF 
onto the IBM-2 deformation energy surfaces. 
}
\end{figure}

\begin{figure}

\centering
\includegraphics[width=\linewidth]{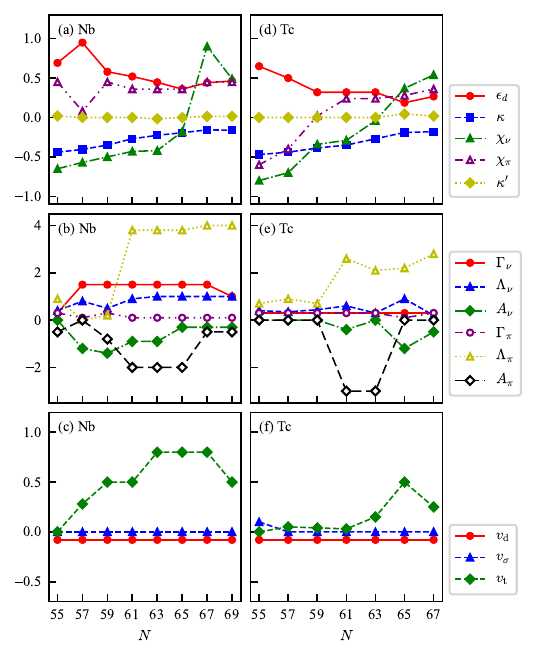}
\caption{\label{fig:para-ibffm}
The IBFFM-2 parameters adopted in the present study 
for the odd-odd $^{96-110}$Nb (left column) 
and $^{98-110}$Tc (right column) nuclei.
}
\end{figure}

\begin{enumerate}
\item 
The constrained SCMF calculations are
performed for the even-even Ge, Zr, Mo, and Ru isotopes 
by means of the relativistic Hartree-Bogoliubov (RHB) 
method \cite{vretenar2005,niksic2011}, using the 
density-dependent point-coupling (DD-PC1) EDF \cite{ddpc1} 
and the separable pairing force 
of finite range \cite{tian2009}. 
The SCMF calculations yield PESs with triaxial quadrupole 
($\beta$ and $\gamma$) shape degrees of freedom. 
The RHB-SCMF PES is then mapped onto the 
expectation value of $\hb$ 
in the boson coherent state \cite{ginocchio1980}, and 
this procedure specifies the 
IBM-2 Hamiltonian parameters \cite{nomura2008,nomura2010}. 
Only the parameter $\kappa'$ is determined separately 
by comparing cranking moments of inertia between 
fermionic and bosonic systems \cite{nomura2011rot}. 
In addition, it is assumed that 
$\kappa_\nu=\kappa_\pi=\kappa/2$ for 
the even-even $^{96-110}$Zr nuclei 
and $\kappa_\nu=\kappa_\pi=0$ MeV for all the other nuclei, 
for simplicity. 

\item 
The single-fermion Hamiltonian 
$\hf$ (\ref{eq:hf}) and boson-fermion interactions 
$\hbf$ (\ref{eq:hbf}) are constructed by using the 
procedure of Refs.~\cite{nomura2016odd,nomura2019dodd}: 
The RHB-SCMF calculations are 
performed for the neighboring odd-$N$ or odd-$Z$ nucleus 
with the constraint on zero deformation to provide quasiparticle 
energies, $\tilde{\epsilon}_{j_\rho}$, and occupation 
probabilities, $v^2_{j_\rho}$, at the spherical configuration 
for the odd nucleon at orbitals $j_\rho$. 
These quantities are then input to $\hf$ and $\hbf$, 
respectively. The remaining three coupling 
constants, $\Gamma_\rho$, $\Lambda_\rho$, and $A_\rho$, 
are determined to fit the 
experimental data for a few low-lying positive-parity 
levels of each odd-$N$ and odd-$Z$ nuclei. 

\item 
The parameters $\Gamma_\rho$, $\Lambda_\rho$, and 
$A_\rho$, which are determined in the previous 
step for the neighboring odd-$N$ and odd-$Z$ nuclei, 
are used for the odd-odd nucleus, in order to reduce 
the number of parameters. 
The quasiparticle energies $\tilde{\epsilon}_{j_\rho}$ 
and occupation probabilities $v_{j_\rho}^2$ 
are, however, newly computed for the odd-odd nucleus 
in the same way as in the previous step, that is, 
by the SCMF calculations constrained to zero deformation, 
since it is more realistic to use these values 
specifically calculated for the odd-odd nucleus 
than to reuse the corresponding values for the 
neighboring odd-mass nuclei.
Then, the parameters of $\hff$ (\ref{eq:hff}) 
are fixed to reproduce, 
to a certain accuracy, the observed 
low-lying positive-parity 
levels of each odd-odd nucleus. 
\end{enumerate}
The strength parameters for the IBFFM-2 
determined by the above procedure for the 
even-even Ge and odd-odd As nuclei are found in Table~I 
of Ref.~\cite{nomura2022beta-ge}. 
While the parameters for 
the even-even Zr and Mo and odd-odd Tc and Nb isotopes 
with the neutron numbers $54 \leqslant N \leqslant 64$ 
were already presented in Ref.~\cite{nomura2024beta} 
(Fig.~3 and Table~II), 
since we extend the calculations beyond $N=64$, 
we show in Figs.~\ref{fig:para-ibm} and 
\ref{fig:para-ibffm} the IBM-2 and IBFFM-2 
parameters for all the considered 
Zr, Mo, Nb, and Tc nuclei, 
for the sake of completeness.

%
%

%
%

%
%
\begin{figure*}
\centering
\includegraphics[width=\linewidth]{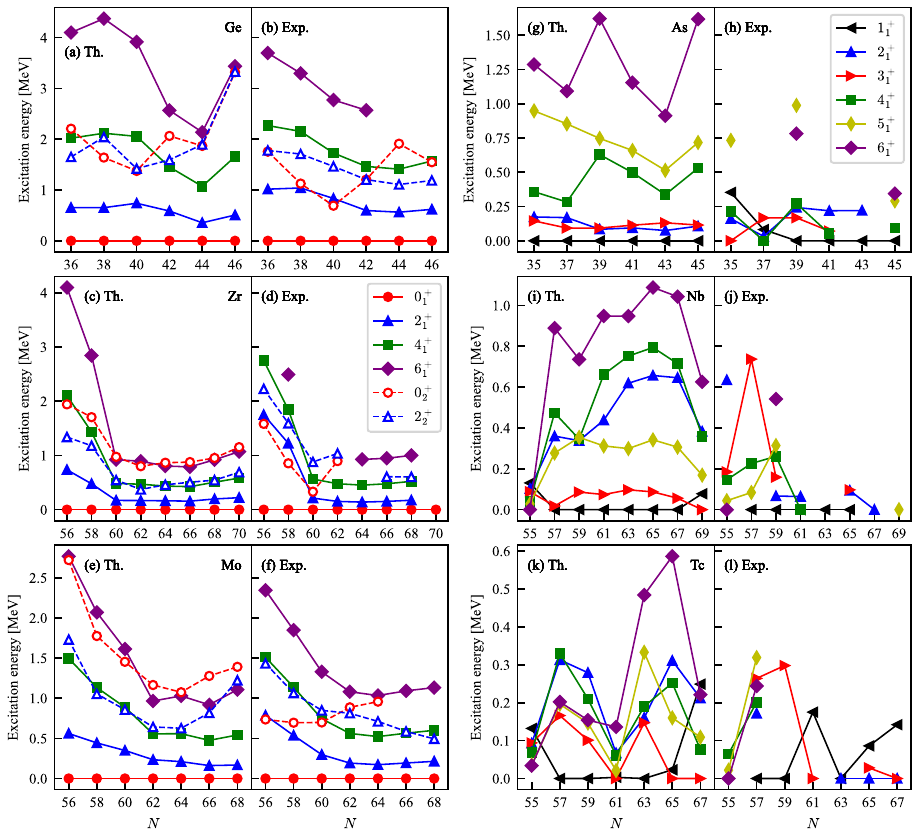}
\caption{\label{fig:energy}
Excitation energies of the even-even $^{68-78}$Ge, 
$^{96-110}$Zr and $^{98-110}$Mo nuclei (first and second columns), 
and odd-odd $^{68-78}$As, $^{96-110}$Nb 
and $^{98-110}$Tc nuclei (third and fourth columns) 
as functions of the neutron number $N$.  
Calculated energies are plotted 
in the first and third columns, and are 
compared with the experimental data \cite{data}, 
shown in the second and fourth columns.}
\end{figure*}

\subsection{Calculated low-energy spectra}

Figure~\ref{fig:energy} depicts 
the calculated excitation energies for low-lying states of 
the even-even nuclei $^{68-78}$Ge, $^{96-110}$Zr,
and $^{98-110}$Mo, and odd-odd nuclei $^{68-78}$As, 
$^{96-110}$Nb, and $^{98-110}$Tc, obtained from the 
diagonalizations of 
the IBM-2 and IBFFM-2 Hamiltonians, 
respectively. 
The experimental data, 
available in the NNDC database \cite{data}, 
are also included in the plots. 
The calculated excitation energies 
for the even-mass Ge and As, 
shown in Figs.~\ref{fig:energy}(a) 
and \ref{fig:energy}(g), are taken from 
those in Figs.~3 and 4 of Ref.~\cite{nomura2022beta-ge}, 
without any modification. 
The predicted energies for the even-even Zr and Mo, 
and odd-odd Nb and Tc isotopes with neutron number up to $N=64$, 
shown in Figs.~\ref{fig:energy}(c), \ref{fig:energy}(e), 
\ref{fig:energy}(i), and \ref{fig:energy}(k), are taken 
from Figs.~2 and 5 of Ref.~\cite{nomura2024beta}. 
An update to the previous studies \cite{nomura2022beta-ge,nomura2024beta} 
is therefore the inclusion of 
the results for those Zr, Mo, Nb, and Tc 
isotopes with $N \geqslant 66$. 
In addition, the calculated $6^+_1$ energy levels 
for both the even-even and odd-odd nuclei are here 
included, which were not considered 
in Refs.~\cite{nomura2022beta-ge,nomura2024beta}. 

What is particularly noteworthy in Fig.~\ref{fig:energy} 
is a drastic change of calculated energy levels for the 
neutron-rich Zr and Mo isotopes near $N=60$, 
indicating a quantum phase transition (QPT) 
in nuclear shapes \cite{cejnar2010}. 
For $N>60$ the predicted ground-state yrast levels, i.e., 
$2^+_1$, $4^+_1$, and $6^+_1$, for the Zr and Mo nuclei 
become compressed in energy, resembling a rotational band. 
This low-lying structure appears to continue toward $N=70$. 
The mapped IBM-2 reproduces this trend of the yrast levels well, 
but overestimates the non-yrast $0^+_2$ and $2^+_2$ levels 
for $N\geqslant 60$. 
For instance, while the measured $2^+_2$ level for the even-even Mo 
isotopes keeps decreasing with $N$, which may 
indicate a shape coexistence 
effect or pronounced 
$\gamma$-softness, the mapped IBM-2 
rather suggests an increase of this level for $N>64$ 
[see Fig.~\ref{fig:energy}(e)]. 
In general, the overestimation of the non-yrast 
levels in the mapped IBM-2 framework has been explained 
by the too large quadrupole-quadrupole 
interaction strength $\kappa$, which is derived from the 
EDF-to-IBM PES mapping, or by the fact that the calculation 
does not include configuration mixing of normal and 
intruder states \cite{duval1981}, which are 
related to coexisting mean-field minima on the PES.

As for the odd-odd nuclei, especially the neutron-rich Nb and Tc, 
it is difficult to determine 
the IBFFM-2 strength parameters, 
as the experimental data for the low-energy spectra are 
scarce. 
Here they are determined mainly to reproduce the spin 
of the ground state, or extrapolated from the neighboring odd-odd 
nuclei where spectroscopic data are available. 
As one can see 
in Figs.~\ref{fig:energy}(i)--\ref{fig:energy}(l),  
there appear notable changes in the calculated 
low-energy level structure 
near the phase-transitional region $N \approx 60$, 
e.g., from $^{96}$Nb ($N=55$) to $^{104}$Nb ($N=63$), 
and from $^{98}$Tc ($N=55$) to $^{106}$Tc ($N=63$). 
These behaviors of low-energy levels in the 
odd-odd systems reflect, to a large extent, 
the structural 
evolution in the adjacent even-even nuclei.

We do not, however, go into further details about 
the nuclear structure aspects 
of these nuclei, since the scope of the 
present paper is rather to investigate their $\beta$-decay properties. 
More thorough discussions about the calculated excitation 
energies, including the comparisons with experimental data in the 
context of the shape QPT and coexistence and possible remedies 
of the deficiencies of the model descriptions, can be found 
in Refs.~\cite{nomura2022beta-ge,nomura2024beta}.

%
%
\begin{figure}[ht]
\centering
\includegraphics[width=\linewidth]{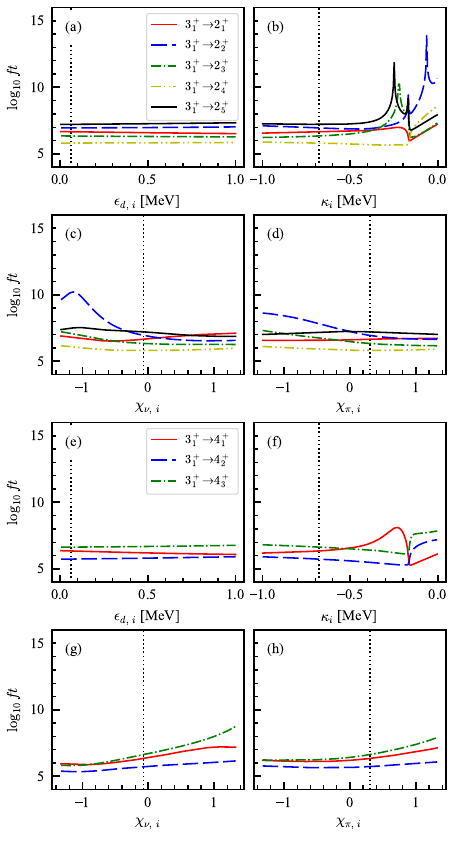}
\caption{Predicted $\ft$ values for the $^{68}$As $\beta$ decay 
as functions of the boson-core parameters for the odd-odd 
parent $^{68}$As nucleus. The vertical dotted line in 
each panel indicates the value of the parameter obtained from 
the RHB-to-IBM mapping procedure.}
\label{fig:ft-as-core}
\end{figure}

%
%
\begin{figure*}[ht]
\begin{center}
\includegraphics[width=0.75\linewidth]{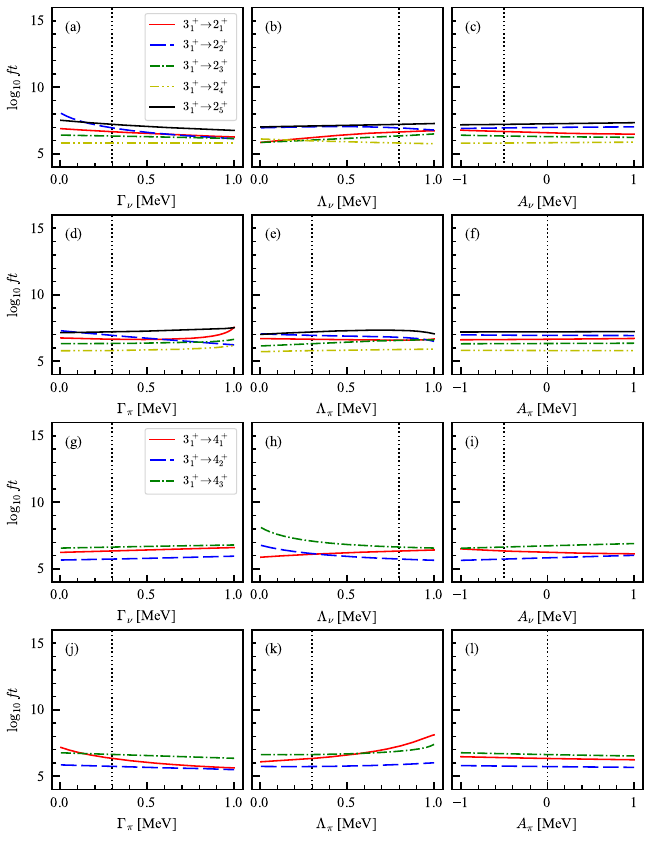}
\caption{Same as Fig.~\ref{fig:ft-as-core}, but 
as functions of the interaction strengths between the 
odd nucleon and even-even boson core for the odd-odd parent $^{68}$As nucleus.}
\label{fig:ft-bfint}
\end{center}
\end{figure*}

%
%
\begin{figure}[ht]
\begin{center}
\includegraphics[width=\linewidth]{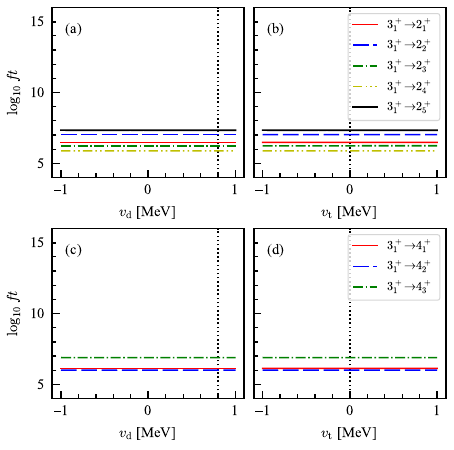}
\caption{Same as Fig.~\ref{fig:ft-as-core}, but 
as functions of the strengths parameters for the 
residual neutron-proton interactions for the 
odd-odd parent $^{68}$As nucleus.}
\label{fig:ft-vt}
\end{center}
\end{figure}

%
%
\begin{figure}[ht]
\begin{center}
\includegraphics[width=\linewidth]{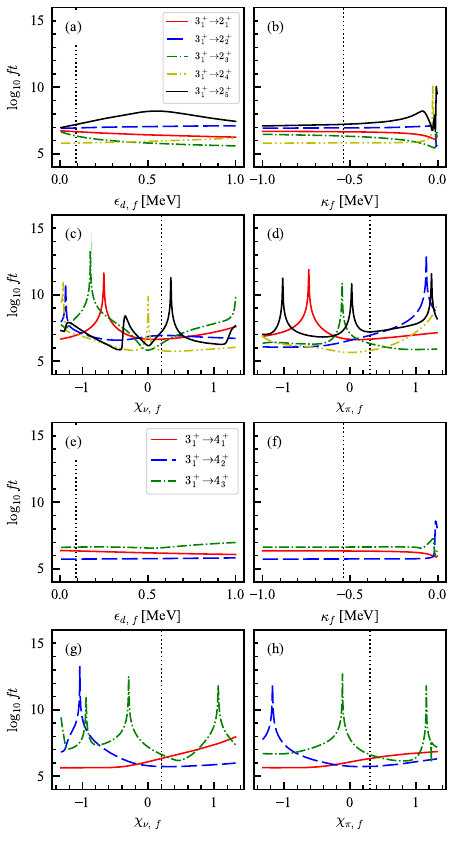}
\caption{Same as Fig.~\ref{fig:ft-as-core}, but 
as functions of the IBM-2 parameters for 
the even-even daughter $^{68}$Ge nucleus.}
\label{fig:ft-ge}
\end{center}
\end{figure}

%
%
\begin{figure*}[ht]
\begin{center}
\includegraphics[width=0.75\linewidth]{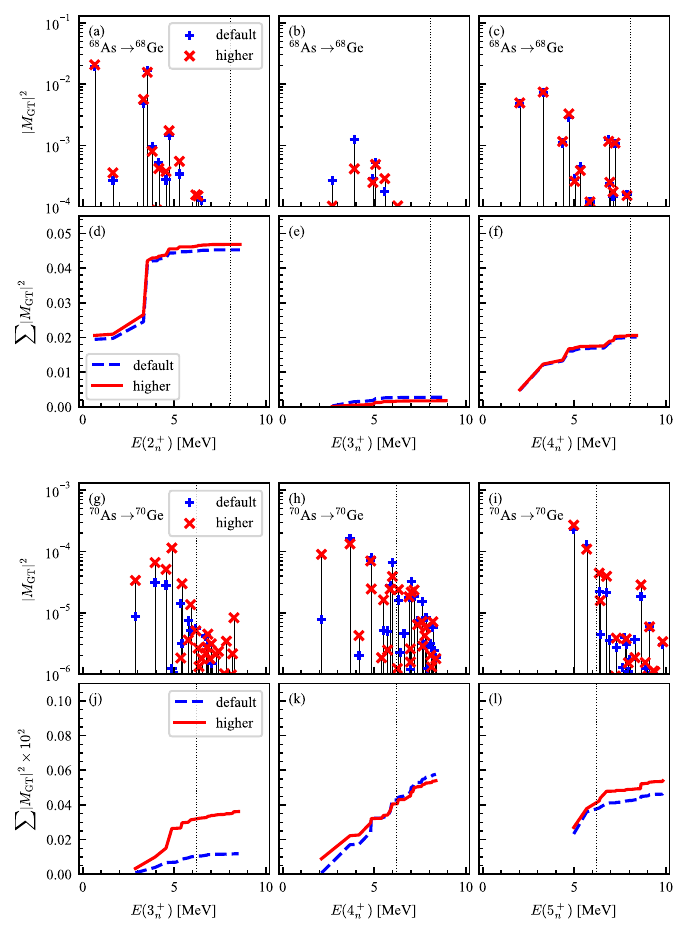}
\caption{Absolute squares of the calculated GT transition 
matrix elements, $\abs{\mgt}^2$, and their running sums 
for the $^{68}$As$(3^+_1)$ and $^{70}$As$(4^+_1)$ 
$\btp$ decays as functions of 
the excitation energies $E(J^\pi_n)$ 
of all those final states with spin parity $J^\pi$ 
obtained from the IBM-2. 
Results of two sets of the IBM-2 calculations, 
which do (``higher'') and do not (``default'') 
include the higher-order terms in the 
one-nucleon transfer operators, are shown. 
The vertical dotted line depicted in each panel 
stands for the experimental $\beta$-decay $Q$-value \cite{data}. 
}
\label{fig:gt-dist-asge}
\end{center}
\end{figure*}

%
%
\begin{figure}[ht]
\begin{center}
\includegraphics[width=\linewidth]{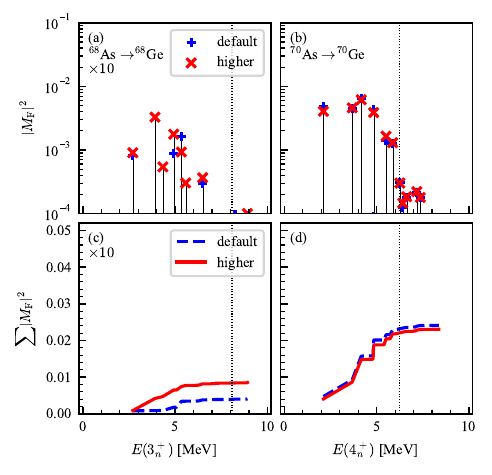}
\caption{Same as Fig.~\ref{fig:gt-dist-asge}, 
but the calculated Fermi 
transition matrix elements, $\abs{\mf}^2$.}
\label{fig:f-dist-asge}
\end{center}
\end{figure}

%
%
\begin{figure}[ht]
\begin{center}
\includegraphics[width=\linewidth]{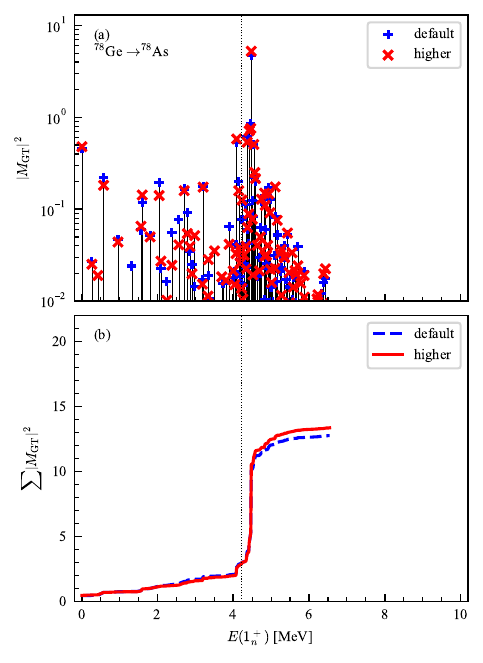}
\caption{Same as Fig.~\ref{fig:gt-dist-asge}, 
but for the $^{78}$Ge$(0^+_1)$ $\btm$ decays 
as functions of the excitation energies  
$E(1^+_n)$ of the $1^+$ final states 
obtained from the IBFFM-2.}
\label{fig:gt-dist-geas}
\end{center}
\end{figure}

%
%
\begin{figure*}[ht]
\begin{center}
\includegraphics[width=\linewidth]{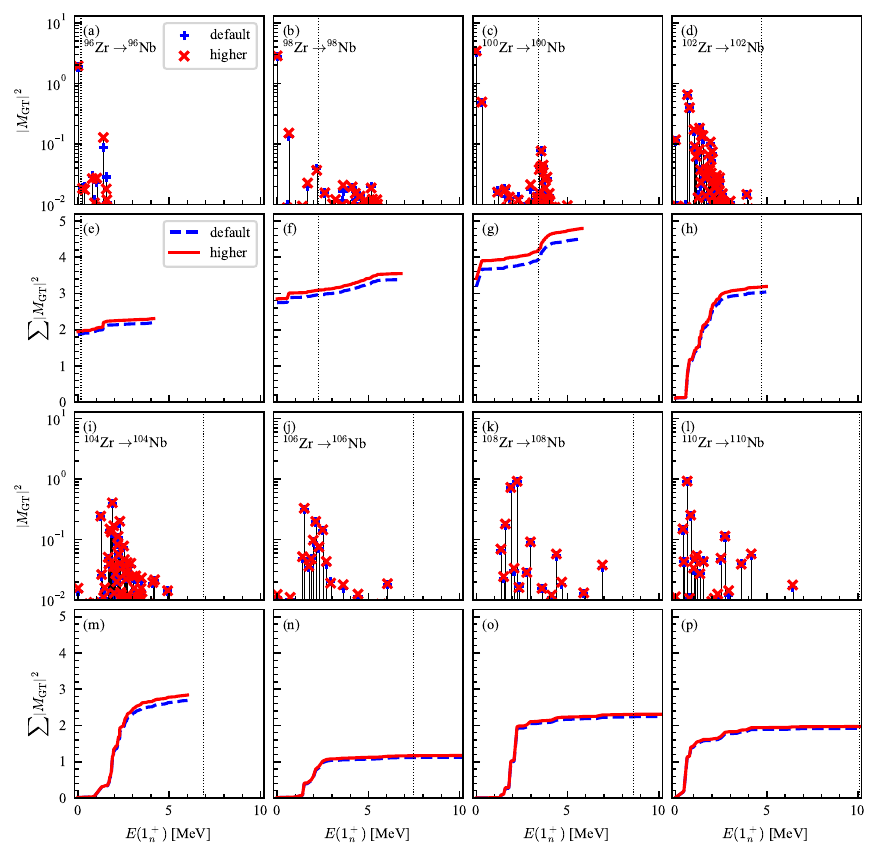}
\caption{Same as the Fig.~\ref{fig:gt-dist-geas} but for the 
$^{96-110}$Zr$(0^+_1)$ $\btm$ decays.
}
\label{fig:gt-dist-zrnb}
\end{center}
\end{figure*}

%
%
\begin{figure*}[ht]
    \begin{center}
    \includegraphics[width=\linewidth]{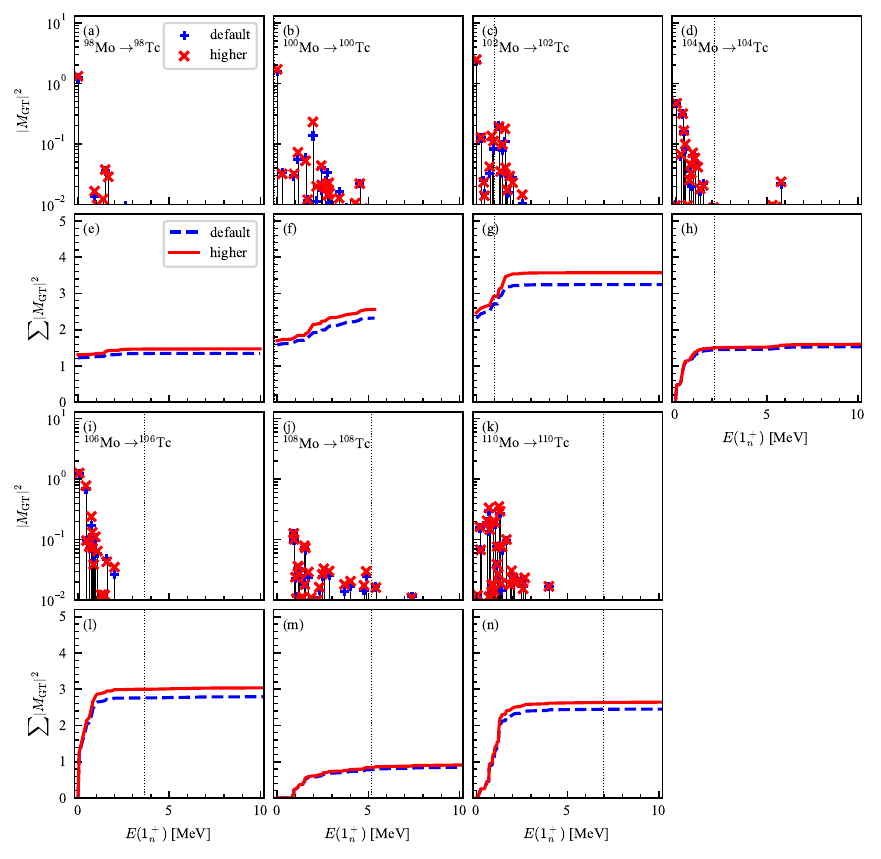}
    \caption{Same as the Fig.~\ref{fig:gt-dist-geas} but for 
    $^{98-110}$Mo$(0^+_1)$ $\btm$ decays.
    }
    \label{fig:gt-dist-motc}
    \end{center}
\end{figure*}

\section{$\beta$ decay\label{sec:beta}}

\subsection{Gamow-Teller and Fermi operators \label{sec:gt-f-op}}

The Gamow-Teller $\hat{T}^{\rm GT}$ 
and Fermi $\hat{T}^{\rm F}$ 
transition operators are here defined by 
\begin{align}
\label{eq:ogt}
\hat{T}^{\rm GT}
&=\sum_{j_{\nu}j_{\pi}}
\eta_{j_{\nu}j_{\pi}}^{\mathrm{GT}}
\left(\hat P_{j_{\nu}}\times\hat P_{j_{\pi}}\right)^{(1)} \; , \\ 
\label{eq:of}
\hat{T}^{\rm F}
&=\sum_{j_{\nu}j_{\pi}}
\eta_{j_{\nu}j_{\pi}}^{\mathrm{F}}
\left(\hat P_{j_{\nu}}\times\hat P_{j_{\pi}}\right)^{(0)} \; ,
\end{align}
respectively, with the coefficients $\eta$ being 
\begin{align}
\label{eq:etagt}
\eta_{j_{\nu}j_{\pi}}^{\mathrm{GT}}
&= - \frac{1}{\sqrt{3}}
\left\langle
\ell_{\nu}\frac{1}{2}j_{\nu}
\bigg\|{\bm\sigma}\bigg\|
\ell_{\pi}\frac{1}{2}j_{\pi}
\right\rangle
\delta_{\ell_{\nu}\ell_{\pi}} \; , \\
\label{eq:etaf}
\eta_{j_{\nu}j_{\pi}}^{\mathrm{F}}
&= - \sqrt{2j_\nu + 1}
\delta_{j_{\nu}j_{\pi}} \; .
\end{align}
$\hat P_{j_\rho}$ in Eqs.~(\ref{eq:ogt}) 
and (\ref{eq:of}) is identified as one of the 
one-particle creation operators
\begin{align}
\label{eq:creation1}
A_{j_\rho m_{j_\rho}}^{\+}
&=\zeta_{j_\rho} a_{{j_\rho}m_{j_{\rho}}}^{\+}
 + \sum_{j_\rho'} \zeta_{j_\rho j_\rho'} 
s_\rho^{\+} (\tilde{d}_{\rho}\times a_{j_\rho'}^{\+})^{(j_\rho)}_{m_{j_{\rho}}} \; , \\
\label{eq:creation2}
B_{j_\rho m_{j_\rho}}^{\+}
&=\theta_{j_{\rho}} s_\rho^{\+} \tilde{a}_{j_{\rho}m_{j_{\rho}}}
 + \sum_{j_{\rho}'} \theta_{j_{\rho}j_{\rho}'} 
(d_{\rho}^{\+}\times\tilde{a}_{j_{\rho}'})^{(j_{\rho})}_{m_{j_{\rho}}} \; ,
\end{align}
and the annihilation operators 
\begin{align}
    \label{eq:annihilation1}
    \tilde A_{j_\rho m_{j_\rho}}
    &= (-1)^{j_\rho - m_{j_\rho}}A_{j_\rho -m_{j_\rho}} \; , \\
    \label{eq:annihilation2}
    \tilde B_{j_\rho m_{j_\rho}}^{\+}
    &= (-1)^{j_\rho - m_{j_\rho}}B_{j_\rho -m_{j_\rho}} \; .
\end{align}
Note that the operators in Eqs.~(\ref{eq:creation1}) and 
(\ref{eq:annihilation1}) conserve the boson number, whereas 
those in Eqs.~(\ref{eq:creation2}) and (\ref{eq:annihilation2}) 
do not. 
The $\hat{T}^{\mathrm{GT}}$ and $\hat{T}^{\mathrm{F}}$ 
operators are constructed as an appropriate combination 
of two of the operators 
in Eqs.~(\ref{eq:creation1})--(\ref{eq:annihilation2}), 
according to the type of the $\beta$ decay under study (i.e., 
$\btp$ or $\btp$) and on the particle or hole nature of 
bosons as well as odd nucleons. 
To be specific,  
$\hat P_{j_{\rho}}$'s considered in the present 
for the different $\beta$-decay processes 
are as follows:
\begin{align}
\label{eq:opt-1}
 \hat P_{j_{\nu}} = B^\+_{j_{\nu}m_{\nu}}
\quad , \quad
\hat P_{j_{\pi}} = \tilde A_{j_{\pi}m_{\pi}}
\end{align}
for the $^{68,70}$As$\to^{68,70}$Ge $\btp$ decays, 
\begin{align}
\label{eq:opt-2}
\hat P_{j_{\nu}} = A^{\+}_{j_{\nu}m_{\nu}}
\quad , \quad
\hat P_{j_{\pi}} = A^{\+}_{j_{\pi}m_{\pi}}
\end{align}
for the $^{78}$Ge$\to^{78}$As $\btm$ decay, 
\begin{align}
\label{eq:opt-3}
\hat P_{j_{\nu}} = \tilde B_{j_{\nu}m_{\nu}}
\quad , \quad
\hat P_{j_{\pi}} = \tilde B_{j_{\pi}m_{\pi}}
\end{align}
for the $^{96-106}$Zr$\to^{96-106}$Nb and 
$^{98-108}$Mo$\to^{98-108}$Tc $\btm$ decays, and 
\begin{align}
\label{eq:opt-4}
\hat P_{j_{\nu}} = A^{\+}_{j_{\nu}m_{\nu}}
\quad , \quad
\hat P_{j_{\pi}} = \tilde B_{j_{\pi}m_{\pi}}
\end{align}
for the $^{108,110}$Zr$\to^{108,110}$Nb and 
$^{110}$Mo$\to^{110}$Tc $\btm$ decays.  
It is also noted that 
the expressions in Eqs.~(\ref{eq:creation1})--(\ref{eq:annihilation2}) 
are of simplified forms of the most general one-particle 
transfer operators in the boson representation.

Within the generalized seniority scheme 
\cite{dellagiacoma1989,IBFM-Book,dellagiacoma1988phdthesis}, 
the coefficients 
$\zeta_j$, $\zeta_{jj'}$, $\theta_j$ and $\theta_{jj'}$ 
in Eqs.~(\ref{eq:creation1}) and (\ref{eq:creation2}) can 
be given by 
\begin{eqnarray}
    \zeta_{j_\rho} &=& u_{j_\rho} \frac{1}{K'_{j_\rho}} \; , \label{eq:zeta1} \\
    \zeta_{j_\rho j'_\rho} &=& -v_{j_\rho}\beta_{j'_\rho j_\rho}\sqrt{\frac{10}{N_\rho (2j_\rho + 1)}} \frac{1}{KK'_{j_\rho}} \; , \label{eq:zeta2} \\
    \theta_{j_\rho} &=& \frac{v_{j_\rho}}{\sqrt{N_\rho}} \frac{1}{K''_{j_\rho}} \; , \label{eq:theta1} \\
    \theta_{j_\rho j'_\rho} &=& u_{j_\rho}\beta{j'_\rho j_\rho}\sqrt{\frac{10}{2j_\rho + 1}} \frac{1}{KK''_{j_\rho}} \label{eq:theta2} \; .
\end{eqnarray}

The factors $K$, $K'_{j_\rho}$ and $K''_{j_\rho}$ are 
defined as 
\begin{align}
    K &= \qty(\sum_{j_\rho j'_\rho}\beta_{j_\rho j'_\rho}^2)^{1/2} \; , \label{eq:K1} \\
    K'_{j_\rho} &= \qty[1 + 2\qty(\frac{v_{j_\rho}}{u_{j_\rho}})^2 \frac{\ev{(\hat{n}_{s_\rho}+1)\hat{n}_{d_\rho}}_{0^+_1}}{N_\rho (2j_\rho + 1)} \frac{\sum_{j'_\rho}\beta^2_{j'_\rho j_\rho}}{K^2}]^{1/2} \; , \\
    K''_{j_\rho} &= \qty[\frac{\ev{\hat{n}_{s_\rho}}_{0^+_1}}{N_\rho} + 2\qty(\frac{u_{j_\rho}}{v_{j_\rho}})^2\frac{\ev{\hat{n}_{d_\rho}}_{0^+_1}}{2j_\rho + 1} \frac{\sum_{j'_\rho}\beta^2_{j'_\rho j_\rho}}{K^2}]^{1/2} \; ,
\end{align}
where $\hat{n}_{s_\rho}$ is the number operator for 
$s_\rho$ bosons and $\ev{\cdots}_{0^+_1}$ represents 
the expectation value of a given operator in the $0^+_1$ 
ground state of the even-even nucleus. 
The expressions 
in Eqs.~(\ref{eq:zeta1})--(\ref{eq:theta2}) 
are specified only by the occupation 
$v_{j_\rho}$ and unoccupation $u_{j_\rho}$ amplitudes 
computed by the RHB-SCMF calculations, and 
the same ($u_{j_\rho},v_{j_\rho}$) 
values as those used in the IBFFM-2 
calculations for the odd-odd nuclei are employed. 
No phenomenological parameter 
is introduced in the $\hat T^{\rm GT}$ 
and $\hat T^{\rm F}$ operators.

The $ft$ values are calculated by using the 
Fermi, $\mf = \mel*{I_f}{|\hat T^{\rm F}|}{I_i}$, 
and GT, $\mgt = \mel*{I_f}{|\hat T^{\rm GT}|}{I_i}$, 
reduced matrix elements 
for the transitions between the initial 
state $I_i$ of the parent nucleus 
and the final state $I_f$ 
of the daughter nucleus:
\begin{eqnarray}
 ft = \frac{K}{|\mf|^2+\left(\ga/\gv\right)^2 |\mgt|^2} \; ,
 \label{eq:ft}
\end{eqnarray}
with the factor $K=6163$ (in seconds), 
$\ga=1.27$ and $\gv=1$ being the 
axial-vector and vector coupling constants, 
respectively. 

For a more detailed account on the formalism of the 
$\beta$-decay operators within the IBFFM-2, the reader is 
referred to Refs.~\cite{dellagiacoma1989,IBFM-Book,dellagiacoma1988phdthesis}.

\subsection{Parameter dependence of $\beta$-decay $\ft$ of $^{68}$As\label{sec:as68}}

As mentioned in Sec.~\ref{sec:ham}, 
the independent parameters for the IBM-2 Hamiltonian 
involved in the present study 
are $\epsilon_d$, $\kappa$, 
$\chi_\nu$, $\chi_\pi$, and $\kappa'$.
In a number of microscopic and 
phenomenological IBM-2 calculations 
carried out by now, however, 
it has been shown \cite{IBM} 
that a simplified form of 
the Hamiltonian consisting only of the $\hat n_d$, 
and $\hat Q_\nu \cdot \hat Q_\pi$ terms 
is adequate to describe the low-energy quadrupole 
collective states of most of the medium-heavy and 
heavy nuclei. 
Thus, in the following we investigate dependencies of the 
results on the parameters 
$\epsilon_d$, $\kappa$, $\chi_\nu$, and $\chi_\pi$ 
concerning the bosonic interactions, 
while keeping the parameter $\kappa'$ 
(for the $\hat L\cdot \hat L$ term) unchanged 
for each nucleus. 
To avoid confusions, 
we express from now on those IBM-2 
parameters used for the parent nuclei by 
a subscript $i$, representing the initial 
state, and those for the daughter nuclei 
with a subscript $f$, representing the 
final state. 
Six parameters in the IBFFM-2 Hamiltonian, 
$\Gamma_\nu$, $\Lambda_\nu$, $A_\nu$, 
$\Gamma_\pi$, $\Lambda_\pi$, and $A_\pi$, 
which are the coefficients of the boson-fermion 
interactions, 
and the parameters $\vd$, and $\vt$ in 
the residual interaction $\hff$ 
are to be varied. 
The quasi-particle energies 
$\tilde\epsilon_{\jr}$ in $\hf$, and occupation probabilities 
$v^2_{\jr}$, which appear in the boson-fermion 
interaction $\hbf$, GT and Fermi operators, 
are kept unchanged.

Figures~\ref{fig:ft-as-core}--\ref{fig:ft-ge} show calculated 
$\ft$ values of the $\btp$ decay of the $3^+_1$ 
ground state of $^{68}$As into the lowest five $2^+$ and three 
$4^+$ states of $^{68}$Ge as functions of the Hamiltonian parameters.
From Fig.~\ref{fig:ft-as-core} one can see 
that the $\ft$ values have little dependence on 
$\epsilon_{d,i}$, $\chi_{\nu,i}$ and $\chi_{\pi,i}$, which are IBFFM-2 
parameters for the boson core of $^{68}$As. 
The calculated $\ft$ values are much more sensitive 
to the parameter $\kappa_i$, as they exhibit significant changes 
at $\kappa_i \approx \SI{-0.2}{\MeV}$ for both the 
$3^+_1 \to 2^+$ and $3^+_1 \to 4^+$ decays 
[see Figs.~\ref{fig:ft-as-core}(b) and \ref{fig:ft-as-core}(f)]. 
The result is consistent with that obtained in 
our previous study on neutron-rich Zr $\btm$ 
decays \cite{homma2024}, in which the $\ft$ values 
were shown to be particularly 
sensitive to the quadrupole-quadrupole interaction strength 
$\kappa$ in odd-odd nuclei. 

As shown in 
Fig.~\ref{fig:ft-bfint}, 
the calculated $\ft$ values are stable against variations 
of the boson-fermion parameters. 
A similar conclusion was reached in Ref.~\cite{homma2024}.

Figure~\ref{fig:ft-vt} indicates that the predicted $\ft$ values 
are constant when one varies the IBFFM-2 parameters 
$v_\mathrm{t}$ and $v_\mathrm{d}$, which are 
interaction strengths between unpaired nucleons 
of the odd-odd As nuclei. 
This is quite at variance with the finding in Ref.~\cite{homma2024}, 
which showed a strong dependence 
of the $\ft$ values for the Zr $\btm$ decays on 
the strength parameter for the tensor interaction,  
$v_\mathrm{t}$, of the odd-odd daughter Nb nuclei,

Figure~\ref{fig:ft-ge} shows 
dependencies of the $\ft$ values on the IBM-2 
strength parameters for the daughter nucleus $^{68}$Ge. 
One can see that the $\ft$ values are insensitive to 
the parameter $\epsilon_{d,f}$. 
Significant changes of the $\ft$ values for 
both the $3^+ \to 4^+$ and $3^+ \to 2^+$ decays with 
the quadrupole-quadrupole strength 
parameter $\kappa_{f}$ are observed 
within the range $-0.05 < \kappa_f < 0$ MeV 
[see Figs.~\ref{fig:ft-ge}(b), and \ref{fig:ft-ge}(f)]. 
For those $\kappa_f$ values that are larger in magnitude, 
$|\kappa_f| \gg 0.05$ MeV, the $\ft$ values are more stable 
or only exhibit a gradual change with this parameter.

The $\ft$ values also show some dependence on 
the parameter $\chi_{\nu,f}$, 
for the $3^+_1 \to 2^+_3$ and $3^+_1 \to 2^+_5$ 
transitions in particular [see Fig.~\ref{fig:ft-ge}(c)].
There is also something peculiar that is generally 
observed in the $\ft$ systematic with the $\chi_{\nu,f}$ 
and $\chi_{\pi,f}$ [see Figs.~\ref{fig:ft-ge}(c), \ref{fig:ft-ge}(d), 
\ref{fig:ft-ge}(g), and \ref{fig:ft-ge}(h)]: 
at particular values of these parameters 
anomalously large $\ft$ values are obtained. 
As noted in the previous article \cite{homma2024}, 
such ``spike'' patterns seem to occur rather accidentally, 
since with specific combinations of the IBM-2 as well as 
IBFFM-2 parameters cancellations among many different terms 
in the GT matrix elements occur to a great extent, leading to 
unexpectedly small $|\mgt|^2$ values, hence very large $\ft$ values. 
As shown in Ref.~\cite{homma2024}, however, 
these peaks are not of crucial importance for our purpose, 
especially when we calculate the running sum 
$\left| M_\mathrm{GT} \right|^2$ 
taken over large numbers of states 
of the final nucleus, for which 
the local behaviors of $|\mgt|^2$ strengths 
do not play a role.

A major conclusion drawn from the present analysis 
is that the predicted $\ft$ values for the 
$\btp$ decays of odd-odd As nuclei 
within the mapped IBM-2 and IBFFM-2 
are especially sensitive to the quadrupole-quadrupole 
interaction strength $\kappa_i$ in the 
parent or odd-odd As nuclei. 
This finding appears to be robust, 
in view of the fact that similar parameter sensitivities of the 
calculated $\ft$ values were observed in our 
previous study of \cite{homma2024} for the 
$\beta$ decays of neutron-rich Zr region 
with mass $A \approx 100$. 
In that study, we analyzed as an illustrative example 
the $\btm$ decays of the even-even $^{96-102}$Zr isotopes, 
and concluded that the predicted $\ft$ values 
depended significantly on the quadrupole-quadrupole 
interaction strength $\kappa$ used for the 
even-even core Hamiltonian for the daughter 
odd-odd Nb isotopes.

\subsection{Impacts of higher order terms on $\beta$-decay properties\label{sec:hoc}}

We now turn to investigate possible impacts of higher-order terms 
in the one-nucleon transfer operators to $\beta$-decay 
properties. 
The operators 
in Eqs.~(\ref{eq:creation1}) and (\ref{eq:creation2}) 
can be extended to include the additional terms
%
%
\begin{align}
    A'^\+_{j_\rho m_\rho} 
&= A^\+_{j_\rho m_\rho} + 
    \sum_{j'_\rho}{\zeta'_{j_\rho j'_\rho}\left[d_{\rho}^\+\times 
    \left(\tilde{s}_\rho\times a^\+_{j'_\rho}\right)\right]^{(j_\rho)}_{m_\rho}}
\nonumber \\
&    + \sum_{j'_\rho j_{\rho}''}{\zeta_{j_\rho j'_\rho j''_\rho}\left[d_{\rho}^\+\times 
    \left(\tilde{d}_\rho\times a^\+_{j''_\rho}\right)^{(j'_\rho)}_{m'_\rho}\right]^{(j_\rho)}_{m_\rho}} \; ,
\label{eq:creation1_high}\\
B'^\+_{j_\rho m_\rho} &= B^\+_{j_\rho m_\rho} + 
\sum_{j'_\rho}{\theta'_{j_\rho j'_\rho}s^\+_\rho \left[\tilde{d}_\rho \times 
\left(s^\+_\rho\times \tilde{a}_{j'_\rho}\right)\right]^{(j_\rho)}_{m_\rho}}
\nonumber \\
&    + \sum_{j'_\rho j_{\rho}''}{\theta_{j_\rho j'_\rho j''_\rho}s^\+_\rho \left[\tilde{d}_\rho \times 
    \left(d^\+_\rho\times \tilde{a}_{j''_\rho}\right)^{(j'_\rho)}_{m'_\rho}\right]^{(j_\rho)}_{m_\rho}}
\label{eq:creation2_high} \; ,
\end{align}
and their conjugate operators
\begin{align}
\label{eq:annihilation1_high}
& \tilde{A'}_{j_{\rho}m_{\rho}} 
= (-1)^{j_{\rho}-m_{\rho}} {A'}_{j_{\rho}-m_{\rho}}
\\
\label{eq:annihilation2_high}
& \tilde{B'}_{j_{\rho}m_{\rho}} 
= (-1)^{j_{\rho}-m_{\rho}} {B'}_{j_{\rho}-m_{\rho}} \; ,
\end{align}
where the first terms on the right-hand 
sides of the above two equations 
were introduced in 
Eqs.~(\ref{eq:creation1})--(\ref{eq:annihilation2}), 
and are hereafter referred to as leading-order terms. 
Under the condition that the one-particle 
transition operator acts only between states that differ 
in generalized seniority quantum numbers by 
$\Delta \tilde{\nu} = \pm 1$, the 
coefficients of the second terms of (\ref{eq:creation1_high}) 
and (\ref{eq:creation2_high}), i.e., 
$\zeta'_{j_\rho j'_\rho}$ and $\theta'_{j_\rho j'_\rho}$, 
should vanish. 
The coefficients of the 
third term of Eq.~(\ref{eq:creation1_high}), 
$\zeta_{j_\rho j'_\rho j''_\rho}$, 
are obtained by equating the matrix elements 
of the operator 
\begin{eqnarray}
 \left[d^\+\times\left(\tilde{d}_\rho\times a^\+_{j'_\rho}\right)^{(j'_\rho)}_{m'_\rho}\right]^{(j_\rho)}_{m_\rho} \; ,
\end{eqnarray} 
to the corresponding ones in the shell-model (fermion) space. 
The coefficients $\theta_{j_\rho j'_\rho j''_\rho}$ are 
here assumed to be equal to 
$\zeta_{j_\rho j'_\rho j''_\rho}$, for the sake of simplicity. 
$\zeta_{j_\rho j'_\rho j''_\rho}$'s are calculated by following 
the procedure of Ref.~\cite{barea2002}: 
\begin{enumerate}
\item 
Construct a set of $n$ fermion states, 
$\ket{F,(i);JM}\ (i = 1,\dots,n)$, and 
the corresponding boson states, 
$\ket{B,(i);JM}\ (i = 1,\dots,n)$, which 
are orthonormal.

\item
Construct the overlap matrix $\Theta^J$ of 
the fermionic states for each angular momentum $J$. 
For the last term of Eq.~(\ref{eq:creation1_high}), 
the matrix elements of $\Theta^J$ are given as 
\begin{align}
\Theta^J_{mr} 
=&\delta_{mr} + 10\left(\sum_{i=1}^{n}
\delta_{j_i,J}\right)
\nonumber\\
&\times\sum_{k=1}^{n}\beta_{j_kj_r}\beta_{j_kj_m}
\qty{\mqty{
j_r & 2 & j_k \\
j_m & 2 & J}
} \; ,
\label{eq:overlap}
\end{align}
where $\beta_{j_1j_2}$'s are defined in Eq.~(\ref{eq:exc}), 
and the curly bracket denotes Wigner's $6-j$ symbol.

\item
Diagonalize the overlap matrix $\Theta^J$
\begin{equation}
\Theta^J C^J = C^J\lambda^J \; . 
\label{eq:diagonalize}
\end{equation}
Here $C^J$ is the $n\times n$ square matrix. 
Its columns are the eigenvectors 
of $\Theta^J$ which are normalized 
in the condition $\sum_k (C^J_{ik})^2 = 1$.
$\lambda^J$ is the diagonal matrix which contains 
eigenvalues of $\Theta^J$.

\item
Calculate the coefficients 
$\zeta_{j_\rho j'_\rho j''_\rho}$ with 
the expression 
\begin{align}
&\zeta_{j_\rho j'_\rho j''_\rho}
 = \sum_{J}(-1)^{J+j_\rho+j'_\rho-j''_\rho}
\nonumber \\
&\times\sqrt{(2J+1)(2j'_\rho+1)}\qty{\mqty{2 & j''_\rho & J \\ 2 & j_\rho & j'_\rho}}\phi^J_{j_\rho j''_\rho}
\label{eq:zeta3} \; ,
\end{align}
with
\begin{align}
&\phi^J_{j_\rho j''_\rho} = 
(-1)^{j_\rho -j''_\rho} u_{j_\rho} \sqrt{\frac{2J+1}{2j_\rho+1}}
\nonumber\\
&\times \qty[\delta_{j_\rho j''_\rho} - \qty{C^J \sqrt{\lambda^J} (C^J)^T}_{k_{j_\rho}k_{j''_\rho}}]
\label{eq:phi} \; .
\end{align}
\end{enumerate}
The details of the above procedure 
are found in Ref.~\cite{barea2002}.

The GT and Fermi operators are calculated 
by substituting the operators ${A'}^{(\dagger)}_{\jr\mr}$ 
(\ref{eq:creation1_high}) and ${B'}^{(\dagger)}_{\jr\mr}$ 
(\ref{eq:creation2_high}) in 
the formulas in Eqs.~(\ref{eq:ogt}) and (\ref{eq:of}). 
Note that, in calculating the GT and Fermi matrix elements, 
terms that are products of more than two $d$-boson creation 
or annihilation operators are omitted, due to limitation 
of the current version of the computer code we have at hand. 
For the $^{68}$As $\btp$ decay, for example, 
the code is unable to compute the matrix elements 
of those terms proportional to 
$s_{\nu}^\+ \tilde d_{\nu} d_{\nu}^\+ s_{\pi} d^\+_{\pi} \tilde a_{j_{\nu}}\tilde a_{j_{\pi}}$, 
$d^\+_{\nu} \tilde d_{\pi} d_{\pi}^\+ \tilde a_{j_{\nu}}\tilde a_{j_{\pi}}$, 
and 
$s_{\nu}^\+ \tilde d_{\nu} d_{\nu}^\+ \tilde d_{\pi} d_{\pi}^\+ \tilde a_{j_{\nu}}\tilde a_{j_{\pi}}$, but allows one to calculate those 
matrix elements that are products of any other terms in 
${B'}^\+_{j_{\nu}m_{\nu}} \tilde{A'}_{j_{\pi}m_{\pi}}$ 
[see Eq.~(\ref{eq:opt-1})]. 
New components in the matrix element 
of ${B'}^\+_{j_{\nu}m_{\nu}} \tilde{A'}_{j_{\pi}m_{\pi}}$ 
that are introduced by the inclusion 
of the higher-order terms are those 
of types $s^{\+}_{\nu} \tilde d_{\pi} d^{\+}_{\pi} \tilde a_{j_{\nu}}\tilde a_{j_{\pi}}$ 
and $s^{\+}_{\nu} \tilde d_{\nu} d^{\+}_{\nu} \tilde a_{j_{\nu}}\tilde a_{j_{\pi}}$. 
Note that the second terms in Eqs.~(\ref{eq:creation1_high})
and (\ref{eq:creation2_high}) are omitted, since the 
coefficients of these terms vanish in the seniority
consideration.

Figure~\ref{fig:gt-dist-asge} shows the GT transition strengths and 
their running sums for $\btp$ decays 
of $^{68}$As$(3^+_1)$ and $^{70}$As$(4^+_1)$ 
as functions of the excitation energy 
of each spin-parity state of the respective 
daughter nuclei. 
One can observe that both the GT strengths $|\mgt|^2$ and running 
sums are not significantly affected 
by the inclusion of the higher-order terms. 
An exception is that 
the converged value of the GT running sum for the 
$^{70}$As$(4^+_1)\to^{70}$Ge$(3^+)$ $\btp$ decay 
is, by approximately 
a factor of 4, larger than in the calculation 
that takes into account up to the leading-order terms 
in the one-nucleon transfer operators 
[see Fig.~\ref{fig:gt-dist-asge}(j)].

Figure~\ref{fig:f-dist-asge} shows 
the Fermi transition strengths $|\mf|^2$ and the running sums 
of the $^{68,70}$As $\btp$ decays. 
An effect of introducing the high-order terms 
is that, for the $^{68}$As decay, 
the running sum $\sum|\mf|^2$ converges to a value 
that is twice as large as that obtained without 
the higher-order terms. 
As shown in Fig.~\ref{fig:gt-dist-asge}(e), 
the GT sum $\sum|\mgt|^2$ for the 
$\btp$ decay of the $3^+_1$ state of $^{68}$As to 
the $3^+$ states of $^{68}$Ge is also small, and is 
smaller by an order of magnitude than 
those for the GT transitions to the $2^+$ and $4^+$ 
states of $^{68}$Ge. 
Therefore, the $\btp$ decay of 
$^{68}$As$(3^+_1)$ appears to be dominated by the 
GT transitions to the even-spin ($2^+$ and $4^+$) states. 
Compared with the dominant GT transitions, 
the enhancement by a factor of 2 
of the Fermi strength $\sum|\mf|^2$ for the 
$3^+_1 \to 3^+$ decays by the 
inclusion of the high-order terms is relatively minor 
in the $^{68}$As allowed decays. 
The higher-order terms, however, may have potential impacts 
on the predictions of processes 
such as the superallowed $0^+ \to 0^+$ $\beta$ decays, 
where the GT transitions are forbidden and which are 
therefore determined only by the Fermi transitions. 
In the case of the $^{70}$As decay, 
the present results show that the higher-order 
terms do not appear to have a significant 
influence on the Fermi transitions.

Figures~\ref{fig:gt-dist-geas}--\ref{fig:gt-dist-motc} show the 
Gamow-Teller transition matrix elements $\left|M_\mathrm{GT}\right|^2$ 
and their running sums for $^{78}$Ge, $^{96-110}$Zr, and $^{98-110}$Mo 
$\btm$ decays, respectively, as functions of the excitation 
energy of $1^+$ states of daughter nuclei. 
In all decays, 
the effects of the higher-order terms on the magnitude of 
$\left|M_\mathrm{GT}\right|^2$ are small, within a few percent at most. 
However, there is a general trend that, for 
the $0^+_1 \to 1^+$ $\btm$ decays of all the even-even nuclei, 
the GT sums are systematically 
increased by the inclusion of the higher-order terms, 
thus making the decay occur a little faster.

%
%
\begin{figure*}[ht]
\begin{center}
\includegraphics[width=\linewidth]{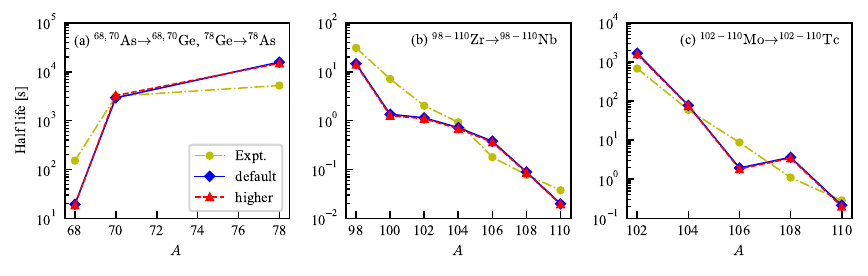}
\caption{
Predicted $\beta$-decay half-lives 
that do (``higher'', red triangles) 
and do not (``default'', blue diamonds) include higher-order terms in 
the one-nucleon transfer operators. 
Available experimental data (``Expt.'', yellow circles) 
are also shown.}
\label{fig:half-life}
\end{center}
\end{figure*}

\subsection{$\beta$-decay half lives\label{sec:half-life}}

Using the GT and Fermi strengths obtained in the 
previous section, we compute the $\beta$-decay half-lives. 
The half-life is defined by the formula
%
%
\begin{equation}
T_{1/2} = \left(\sum_{0\leq E_f \leq Q_\beta}\frac{1}{t_{1/2}^{\mathrm{par}}}\right)^{-1} \; ,
\label{eq:half_life}
\end{equation}
where $Q_\beta$ stands for the $\beta$-decay 
$Q$ value, which we take from experiment \cite{data}, 
and $t_{1/2}^\mathrm{par}$ denotes the partial 
half-life: 
\begin{equation}
    t_{1/2}^\mathrm{par} = \frac{6163}{f(Z,E_f)\left|M_{if}\right|^2} \; .
    \label{eq:par_half_life}
\end{equation}
Here $f(Z,E_f)$ is defined as 
\begin{equation}
    f(Z,E_f) = \int_{1}^{E_f} \dd{E} F(Z,E) p E \left(E_f-E\right)^2 
    \label{eq:f}
\end{equation}
in the units of 
$\hbar = m = c = 1$ \cite{konopinski65}. 
$E_f$ represents the energy difference 
between the ground state of the parent nucleus 
and the excited state of the daughter nucleus $E_x$, i.e., 
$E_f = Q_\beta - E_x$. 
In Eq.~(\ref{eq:f}), momentum $p$ 
and Fermi function $F(Z,E)$ are given as 
\begin{align}
& p = \sqrt{E^2-1}
\label{eq:p}\\
& F(Z,E) 
= 2(1+\gamma_0)\left(\frac{2pR}{\hbar}\right)^{-2(1-\gamma_0)}e^{\pi\nu}\left|\frac{\Gamma(\gamma_0+i\nu)}{\Gamma(2\gamma_0+1)}\right|^2 \; ,
\label{eq:fermi_dist}
\end{align}
where
\begin{align}
    \gamma_0 &= \sqrt{1-(\alpha Z)^2} \label{eq:gamma_0} , \\
    \nu &= \pm\frac{\alpha ZE}{cp} \label{eq:nu}\quad (\mathrm{for}\ \beta^\mp\ \mathrm{decay}) , \\
    R &= \frac{\alpha A^{1/3}}{2} \label{eq:radius} \; .
\end{align}
with $\alpha$ being 
the fine structure constant, $\alpha=1/137$.

By using the above formulas we compute the half-lives 
of the $\btp$ decays of $^{68,70}$As, and the $\btm$ decays of 
$^{78}$Ge, $^{98-110}$Zr, and $^{102-110}$Mo. 
Note, however, that we do not calculate the 
half-lives of $^{76}$Ge, $^{96}$Zr, and $^{98,100}$Mo $\btm$ 
decays, since they are stable nuclei, which in general 
do not have states satisfying the condition 
$0\le E_f \le Q_\beta$ in Eq.~(\ref{eq:half_life}) as 
the corresponding $Q$ values are negative. 
The $^{72,74}$As nuclei are unstable, and their observed 
ground states have spin and parity $2^-$. 
However, since the present version of the computer code 
handles only the allowed (GT and Fermi) transitions, 
we only consider positive-parity states of these nuclei 
and their decays to the states of the even-even 
daughter $^{72,74}$Ge of the same parity. 
To compute the half-lives, 
we use the GT and Fermi transition strengths, 
$\left|M_\mathrm{GT}\right|^2$ and $\left|M_\mathrm{F}\right|^2$, 
in both cases where higher-order terms are included in the 
corresponding one-nucleon transfer operators, and where 
these additional terms are not taken into account.

Figure~\ref{fig:half-life} compares the predicted 
half-lives in the mapped IBM-2 calculations that do and do not 
include the higher-order terms with the 
experimental data. 
Also, Table~\ref{tab:half-life} summarizes 
all the numerical values that are plotted 
in Fig.~\ref{fig:half-life}, 
since it is especially hard to distinguish between 
the predicted values by the two sets of the 
IBM-2 calculations in the figure. 
As one can see from Fig.~\ref{fig:half-life}, 
overall the calculated results 
show a clear trend that half-lives become shorter as one moves away from 
the $\beta$-stability line to the neutron-rich side, 
and the order of magnitude of the calculated half-lives 
is also roughly consistent with data. 
In addition, the inclusion of the 
higher-order terms in the one-nucleon transfer 
operators is rather minor, as it changes the half-lives 
by approximately 5-10 \% (see, Table~\ref{tab:half-life}). 
One cannot, therefore, firmly conclude as to whether or not 
the higher-order term effects are significant enough 
to improve description of the $\beta$-decay half-life calculations 
within the present implementation of the IBM.

We now have a closer look into Fig.~\ref{fig:half-life}. 
In Table~\ref{fig:half-life}(a), the half-lives of 
$^{68}$As$(3^+_1)\to^{68}$Ge obtained from the 
IBM-2 calculations are, by an order of magnitude,
smaller than experimental data, while the calculated 
half-lives of $^{70}$As$(4^+_1)\to^{70}$Ge are quite close to the 
experimental value. The calculated half-lives of the 
$^{78}$Ge $\btm$ decay are by approximately a factor of 3 
larger than the experimental value.
Figure~\ref{fig:half-life}(b) shows that 
the IBM-2 calculations generally underestimate the half-lives 
of Zr isotopes with $N \leqslant 62$. 
For the $\btm$-decay half-lives of the neighboring 
$^{102-110}$Mo nuclei, shown in Fig.~\ref{fig:half-life}(c), 
the present calculation reproduces the observed 
decreasing pattern of half-lives with the neutron number reasonably well. 
A local irregularity or increase of the calculated 
half-life from $^{106}$Mo to $^{108}$Mo may reflect 
the fact that they are close to the middle of the 
neutron major shell $N=66$, after which bosons in the IBM-2 
are treated as holes, and therefore 
the number of (like-hole) bosons 
turns out to decrease for those neutron numbers with $N\geqslant 66$. 
It should be noted, nevertheless, 
that the calculated half-lives for both the 
$^{106}$Mo to $^{108}$Mo $\btm$ decays are in the same 
order of magnitude as the corresponding 
experimental values.

%
%
\begin{table}[htb!]
\caption{\label{tab:half-life}
Calculated and experimental \cite{data} 
$\beta$-decay half-lives (in seconds) 
for those nuclei plotted in Fig.~\ref{fig:half-life}. 
The calculated results within the mapped IBM-2 
that do (``Higher'') and do not (``Default'') include 
higher-order terms in the one-nucleon transfer 
operators are compared. 
}
\begin{center}
\begin{ruledtabular}
\begin{tabular}{ccccc}
\multirow{2}{*}{Parent} &
\multirow{2}{*}{Daughter} &
\multirow{2}{*}{Expt.} &
\multicolumn{2}{c}{IBM-2} \\
\cline{4-5}
& & &
\textrm{Default} &
\textrm{Higher} \\
\hline
$^{68}$As & $^{68}$Ge & 152 & 19 & 19 \\
$^{70}$As & $^{70}$Ge & 3156 & 2927 & 3261 \\
$^{78}$Ge & $^{78}$As & 5220 & 15680 & 14886 \\
$^{98}$Zr & $^{98}$Nb & 30.7 & 14.5 & 14.0 \\
$^{100}$Zr & $^{100}$Nb & 7.1 & 1.3 & 1.323 \\
$^{102}$Zr & $^{102}$Nb & 2.01 & 1.14 & 1.09 \\
$^{104}$Zr & $^{104}$Nb & 0.922 & 0.716 & 0.681 \\
$^{106}$Zr & $^{104}$Nb & 0.178 & 0.378 & 0.361 \\
$^{108}$Zr & $^{108}$Nb & 0.0774 & 0.0886 & 0.0859 \\
$^{110}$Zr & $^{110}$Nb & 0.0376 & 0.0199 & 0.0194 \\
$^{102}$Mo & $^{102}$Tc & 678 & 1673 & 1567 \\
$^{104}$Mo & $^{104}$Tc & 59.4 & 77.2 & 75.8 \\
$^{106}$Mo & $^{106}$Tc & 8.73 & 1.94 & 1.82 \\
$^{108}$Mo & $^{108}$Tc & 1.106 & 3.631 & 3.471 \\
$^{110}$Mo & $^{110}$Tc & 0.287 & 0.216 & 0.201 \\
\end{tabular}
\end{ruledtabular}
\end{center}
\end{table}

\section{Summary and conclusions\label{sec:summary}}

In summary, we have analyzed  
parameter sensitivities of the 
mapped IBM-2 and IBFFM-2 descriptions 
of the $\beta$-decay properties of 
neutron-deficient Ge and As isotopes, 
and neutron-rich Zr and Mo isotopes. 
Within this framework the strength parameters for the 
IBM-2 Hamiltonian describing even-even nuclei, 
single-particle energies and occupation probabilities 
of unpaired nucleons, 
necessary to build the IBFFM-2 Hamiltonian and 
GT and Fermi transition operators, 
were completely determined by using the 
results of the RHB-SCMF calculations 
employing the DD-PC1 EDF and separable pairing interaction. 
Only a few coupling constants of the boson-fermion 
interactions and residual interaction between 
an odd neutron and an odd proton had to be 
adjusted to reproduce the experimental 
low-energy spectra in odd-mass and odd-odd nuclei.

From the behaviors of the calculated $\ft$ values 
as functions of a model parameter, 
it has been found that the $\btp$ decay of $^{68}$As 
depends on the quadrupole-quadrupole interaction 
strength parameter of the parent odd-odd nucleus (i.e., $^{68}$As). 
This finding is consistent with our previous 
study on the $\beta$ decay of neutron-rich 
Zr isotopes \cite{homma2024}, in which 
the quadrupole-quadrupole interaction strength 
parameter of odd-odd (daughter) Nb nuclei 
was shown to play an important role in reproducing 
the $\beta$-decay properties. 
On the other hand, 
at variance with the conclusion in Ref.~\cite{homma2024} 
that the strength parameter for the 
tensor-type residual proton-neutron 
interaction for odd-odd nuclei is also relevant, 
in the present analysis 
the calculated $\btp$-decay $\ft$ values were 
shown to be almost independent of this parameter. 
A common feature of the mapped IBM-2 description 
for both the $\btp$ decay of $^{68}$As and $\btm$ decay 
of neutron-rich Zr nuclei is, therefore, that the 
calculated $\ft$ values are most sensitive to 
the quadrupole-quadrupole interaction strength for the 
even-even boson core of an odd-odd nucleus, 
regardless of 
whether it is a parent or daughter nucleus 
of the decay process.

We further investigated effects of 
higher-order terms in the one-nucleon transfer 
operators, which are used to calculate the Gamow-Teller 
and Fermi operators, on the $\beta$-decay 
properties of As, Ge, Zr and Mo isotopes. 
It was found that the inclusion of the 
higher-order terms makes non-negligible contributions to 
the GT and Fermi strength distributions, their running sums, 
and half-lives, but does not alter the overall features 
of these quantities, which are characteristic of the allowed decays. 
The higher-order contributions may have potential 
impacts on the predictions of superallowed $\beta$ decays, 
since in these processes the GT transitions 
are forbidden and only the Fermi ones are present. 
In addition, we have presented an application 
of the mapped IBM-2 framework to compute the $\beta$-decay 
half-lives that has rarely been investigated in the literature. 
The results shown in Fig.~\ref{fig:half-life} 
indicate the ability of this framework 
to describe the overall trend and absolute values 
of the observed $\beta$-decay half-lives of neutron-rich nuclei. 
We consider the employed theoretical method 
as well as the reported results to be promising, 
given the fact that the method is largely based on 
the microscopic EDF calculations and thus 
allows for a consistent description 
of nuclear structures and $\beta$ decay, 
with a minimal set of phenomenological parameters.

A possible future work could be to extend 
the present analysis to carry out 
more systematic half-life calculations, 
involving many other neutron-rich heavy nuclei 
that are experimentally of much interest. 
It is also feasible to extend the present parameter 
sensitivity analysis to the $\beta$-decay studies of
odd-mass nuclei, which were not explicitly considered in 
the present and previous \cite{homma2024} studies, 
within the interacting boson-fermion model (IBFM-2). 
It is especially expected that 
the $\beta$-decay properties of odd-mass nuclei 
also depend on the relevant strength parameters, such as 
that for the quadrupole-quadrupole boson interaction 
in the IBFM-2 Hamiltonian. 
In the meantime, since in the current version of the employed 
theoretical method the strength parameters for the 
boson-fermion and residual neutron-proton interactions are 
fitted to experimental low-energy spectra, it is crucial 
to develop a way to derive or constrain 
these parameters 
only by the microscopic EDF calculations. 
Work along these lines is in progress 
and will be reported elsewhere.

\bibliography{refs}

\end{document}